\documentclass[]{emulateapj}

\bibpunct{}{}{;}{}{}{,}
\usepackage{times}
\usepackage{color}
\usepackage{appendix}
\usepackage{tabularx}
\usepackage{array}
\usepackage{graphicx}
\usepackage[normalem]{ulem}

\renewcommand{\theequation}{\thesection.\arabic{equation}}

\newcommand{\sfr}{\hbox{$\psi$}} 
\newcommand{\ssfr}{\hbox{$\psi_{\mathrm S}$}}
\newcommand{\tauv}{\hbox{$\hat{\tau}_{V}$}}
\newcommand{\ha}{\hbox{H$\alpha$}}
\newcommand{\hb}{\hbox{H$\beta$}}
\newcommand{\mstar}{\hbox{$M_\star$}}
\newcommand{\msun}{\hbox{$M_\odot$}}
\newcommand{\oii}{\hbox{[O\,{\sc ii}]}}
\newcommand{\oiii}{\hbox{[O\,{\sc iii}]}}

\newcommand{\aboh}{\hbox{$12+\log\textrm{(O/H)}$}}
\newcolumntype{x}[1]{%
>{\centering\hspace{0pt}}p{#1}}%

\shorttitle{The rise and fall of the star formation histories of blue galaxies}
\shortauthors{Pacifici et al.}

\begin{document}

\title{The rise and fall of the star formation histories of blue galaxies at redshifts $0.2<$ {\lowercase{z}} $<1.4$}

\author{Camilla~Pacifici,\altaffilmark{1,2,3} Susan~A.~Kassin,\altaffilmark{4,5} Benjamin~Weiner,\altaffilmark{6} St\'ephane~Charlot,\altaffilmark{2} and Jonathan~P.~Gardner\altaffilmark{4}}

\altaffiltext{1}{Yonsei University Observatory, Yonsei University, Seoul 120-749, Republic of Korea}
\altaffiltext{2}{UPMC-CNRS, UMR7095, Institut d'Astrophysique de Paris, F-75014, Paris, France}
\altaffiltext{3}{Max Planck Institute for Astronomy, K\"{o}nigstuhl 17, D-69117 Heidelberg, Germany}
\altaffiltext{4}{Astrophysics Science Division, Goddard Space Flight Center, Code 665, Greenbelt, MD 20771}
\altaffiltext{5}{NASA Postdoctoral Program Fellow}
\altaffiltext{6}{Steward Observatory, 933 North Cherry Street, University of Arizona, Tucson, AZ 85721}

\begin{abstract}
Popular cosmological scenarios predict that galaxies form hierarchically from the merger of many progenitors, each with their own unique star formation history (SFH). We use a sophisticated approach to constrain the SFHs of 4517 blue (presumably star-forming) galaxies with spectroscopic redshifts in the range $0.2<z<1.4$ from the All-Wavelength Extended Groth Strip International Survey (AEGIS). This consists in the Bayesian analysis of the observed galaxy spectral energy distributions with a comprehensive library of synthetic spectra assembled using realistic, hierarchical star formation and chemical enrichment histories from cosmological simulations. We constrain the SFH of each galaxy in our sample by comparing the observed fluxes in the {\it B, R, I} and {\it K$_{s}$} bands and rest-frame optical emission-line luminosities with those of one million model spectral energy distributions. We explore the dependence of the resulting SFHs on galaxy stellar mass and redshift. We find that the average SFHs of high-mass galaxies rise and fall in a roughly symmetric {\it bell-shaped} manner, while those of low-mass galaxies rise progressively in time, consistent with the typically stronger activity of star formation in low-mass compared to high-mass galaxies. For galaxies of all masses, the star formation activity rises more rapidly at high than at low redshift. These findings imply that the standard approximation of exponentially declining SFHs widely used to interpret observed galaxy spectral energy distributions may not be appropriate to constrain the physical parameters of star-forming galaxies at intermediate redshifts.
\end{abstract}


\section{Introduction}
\label{sec:intro}

Constraints on the stellar content of galaxies are often derived from multiwavelength observations by assuming that the star formation history (SFH) of an individual galaxy can be approximated by a simple declining exponential function of the form {\small$\exp(-t/\tau)$}, where $t$ is the galaxy age and $\tau$ the star formation timescale. Such ``$\tau$-models" have been used successfully to estimate, for example, the stellar masses of nearby spiral galaxies from fits of rest-frame optical/near-infrared colors (e.g., \cite{bell2001}). Despite this success, an increasing number of  analyses have pointed out the limitations of $\tau$-models, particularly for applications to studies of high-redshift galaxies. This is the case of Lyman-break galaxies at redshifts $z>2$ (\cite{papovich2001}) and blue star-forming galaxies at $z\sim2$ (\cite{shapley2005}), for which the addition of another stellar component on top of a $\tau$-model can change the derived stellar masses by  a factor of several. In fact, the major episodes of star formation in ultraviolet (UV)-luminous Lyman-break galaxies could last only a few hundred million years (\cite{stark2009}). Reddy et al. (2012) also find that, for galaxies at $1.5 < z < 4$, fits of broadband spectral energy distributions (SEDs) using exponentially declining SFHs produce on average star formation rates (SFRs) 5 to 10 times lower than those inferred from more direct estimates based on the combined analysis of UV and mid-infrared emission. They conclude that SFHs in which the SFR rises with time are more appropriate than declining $\tau$-models for these galaxies. This is reinforced by the conclusions of Maraston et al. (2010) and Pforr et al. (2012) that SED fits of galaxies at $z\sim2$ using declining $\tau$-models produce unrealistically low ages, and that exponentially {\em rising} $\tau$-models should be preferred for these galaxies. In fact, simulations of galaxy formation in a hierarchical universe predict complex, stochastic SFHs of individual galaxies, which are not well approximated by simple declining exponential laws (e.g.,  \cite{lee2009,wuyts2009}).

To better characterize the SFHs of galaxies at moderate and high redshift, we require more sophisticated, physically motivated spectral analysis tools. In this Letter, we achieve this goal by appealing to the approach recently proposed by Pacifici et al. (2012) to constrain the SFHs of 4517 galaxies with secure spectroscopic redshifts in the range $0.2<z<1.4$.  Specifically, we fit the observed broadband SEDs and emission-line strengths of these galaxies using a comprehensive library of synthetic spectra. We assemble this library by combining state-of-the-art models of stellar population synthesis, nebular emission, and attenuation by dust with star formation and chemical enrichment histories derived from the semi-analytic post-treatment of a large-scale cosmological simulation. This approach allows us to characterize the SFHs of galaxies in different mass and redshift ranges in a physically motivated and statistically reliable manner, using realistic, hierarchical mass-assembly histories. The resulting best-estimate SFHs differ significantly from simple exponentially declining laws, and the dispersion about the average SFH in any mass and redshift range can be quite large.

The paper is organized as follows. In \S\ref{sec:data}, we present observations of the {\it B-, R-, I-} and {\it K$_{s}$-}band fluxes and rest-frame optical emission-line luminosities of 4517 blue galaxies from the All-Wavelength Extended Groth Strip International Survey (AEGIS, \cite{davis2007}). In \S\ref{sec:model}, we describe the library of synthetic SEDs built to interpret these observations. We report our results in \S\ref{sec:results} and present our conclusions in \S\ref{sec:concl}. Throughout this Letter, we adopt a standard $\Lambda$CDM cosmology with $\Omega_{\mathrm{M}}=0.3$, $\Omega_{\Lambda}=0.7$, $h=0.7$, and a Chabrier initial mass function (\cite{chabrier2003}).

\section{Data}
\label{sec:data}

To explore the SFHs of galaxies in wide ranges of stellar mass and SFR at different cosmic epochs, we appeal to the AEGIS survey. This combines photometric and spectroscopic observations of the Extended Groth Strip at wavelengths over the whole range from X-ray to radio. In this initial analysis, we restrict ourselves to optical and infrared wavelengths. This is because multiband observations at wavelengths spanning from blue-ward the 4000{\AA} break out to the rest-frame {\it I}-band set valuable constraints on a galaxy's SFH, while spectroscopically measured emission lines are required for solid redshift determinations and precise SFR estimates (e.g., \cite{brinchmann2004}).
From the AEGIS catalog, we extract 6246 galaxies with available photometry at {\it B, R, I} (\cite{coil2004}) and {\it K$_{s}$} (\cite{bundy2006}). The resulting sample is magnitude-limited to {\it K$_{s}$}=22.5. From this sample, we select 4517 potentially star-forming galaxies ($U-B < 1.0$), which do not show contamination by an Active Galactic Nucleus (AGN). In particular, we reject the sources detected by {\it Chandra} in addition to those identified as AGNs following the criteria by Kauffmann et al. (2003) (rest-frame optical emission-line ratios), Yan et al. (2011) (rest-frame optical emission-line ratio versus color), and Donley et al. (2007) (slope of the {\it Spitzer} IRAC SED). Since only 2\% of the potentially star-forming galaxies are detected by Chandra, removing this small percentage does not bias our results. For all 4517 galaxies, Keck/DEIMOS spectra are available from the DEEP2 Redshift Survey (\citep{newman2012}). They cover the wavelength range 6500-9100{\AA} at a resolution of 1.4{\AA} full width at half-maximum (FWHM). The accuracy of the redshifts is 30~km~s$^{-1}$. Such resolution allows us also to extract reliable fluxes of {\oii}$\lambda\lambda3726.0,3728.8$, {\hb}, {\oiii}$\lambda5007$, and {\ha} (\cite{weiner2007}), when these lines fall into the observed wavelength range of the spectra (Table~\ref{tab:data}). This set of photometric and spectroscopic data probes a rest-frame wavelength range from $\lambda \sim$ 3300{\AA}--1.8{$\mu$}m at $z=0.2$ to $\lambda \sim$ 1650--9150{\AA} at $z=1.4$. 

\section{Modeling approach}
\label{sec:model}

We use the recently developed spectral analysis tool of Pacifici et al. (2012) to derive optimal constraints on the SFHs and stellar masses of the galaxies in the sample described above. This tool relies on the combination of a large library of physically motivated SFHs from cosmological simulations, with state-of-art models of stellar population synthesis, nebular emission, and attenuation by dust.

\subsection{Library of model spectral energy distributions}
\label{sec:lib}

Following Pacifici et al. (2012), we build a comprehensive library of model star formation and chemical enrichment histories by performing a post-treatment of the Millennium cosmological simulation (\citep{springel2005}) using the semi-analytic models of De Lucia \& Blaizot (2007). For the present analysis of 4 broadband photometric fluxes and a few emission lines, we draw only 100,000 galaxy SFHs from $z=127$ to the present time. By analogy with Pacifici et al. (2012), we broaden the parameter space by redrawing, for each selected SFH, 10 different combinations of the following parameters: redshift of observation (in the range $0.1<z<1.5$); evolutionary stage (see Sections~2.1 and 3.1.2 in \cite{pacifici2012}); {\it current} (i.e., averaged over a period of 10 Myr before the galaxy is observed) specific SFR ($-2<\log(\ssfr/{\mathrm Gyr}^{-1})<1$); and current gas-phase oxygen abundance ($7<\aboh<9.4$).

We generate a library of 1 million galaxy SEDs by combining this library of star formation and chemical enrichment histories with the latest version of the Bruzual \& Charlot (2003) stellar population synthesis models, the galaxy nebular emission model of Charlot \& Longhetti (2001) (based on the photoionization code {\small CLOUDY}; \cite{ferland1996}), and the 2-component dust model of Charlot \& Fall (2000). We take the same probability distribution of the total optical depth of the dust ($\tauv$) as in Pacifici et al. (2012), but with an upper limit of $\tauv=3$ (instead of 4) which is more appropriate for the current sample of blue star-forming galaxies. We also draw randomly the slope of the attenuation curve in the interstellar medium in the range 0.4--1.1.

This procedure allows us to go beyond the typical approach of using simple idealized functions at fixed metallicity to describe the SFHs of galaxies. A potential drawback is that the star formation and chemical enrichment prescriptions are drawn from a specific semi-analytic model. Our resampling of the library, as described above, minimizes the impact of this choice.

\subsection{Fitting procedure}

We use the Bayesian approach detailed in Pacifici et al. (2012) to constrain the stellar masses, SFRs and SFHs of the 4517 galaxies in the observational sample of \S\ref{sec:data} using the above library of one million model SEDs. For each galaxy in the sample, we compute the likelihood that each model in the library reproduces the observed SED. Specifically, we compare the observed and modelled photometric fluxes in the {\it B}, {\it R}, {\it I} and {\it K$_{s}$} bands (observer frame) and the emission-line luminosities of {\oii}, {\hb}, {\oiii} and/or {\ha} (for each flux measurement, we adopt the errors from the AEGIS catalog and impose a minimum error of 10\%). The exact lines fitted depend on the rest-frame wavelength coverage of the spectra, as listed in Table~\ref{tab:data}. Since the observed galaxies have accurate spectroscopic redshifts, we compute likelihoods only for the model galaxies lying within 0.05 of the spectroscopic redshift. In practice, this amounts to fitting each observed galaxy with $\sim71,500$ models. For each observed galaxy, we use the computed model likelihoods to build the probability density functions of stellar mass, \mstar, and SFR, \sfr\ (normalized to the absolute observed {\it K$_{s}$}-band luminosity). We take the best estimates of \mstar\ and \sfr\ to be the median values of these distributions. The typical uncertainty (defined as half the 16th-84th percentile range of the cumulative probability density function) is around 0.1 for \mstar\ and 0.3 for \sfr. Actual average uncertainties in bins of stellar mass and redshift are indicated in parentheses in Table~\ref{tab:data}. For each galaxy in the sample, we also retain the best-fit SFH.

\begin{table*}
\begin{center}
\begin{small}
\caption{Median stellar masses and SFRs of AEGIS galaxies in different redshift bins (average uncertainties on individual measurements in each bin are indicated in parentheses).}
\label{tab:data}
\begin{tabular}{c c c c c c c c c c c}
\hline
&&&&&&&&&&\\
Redshift range & Observed emission lines&\multicolumn{3}{c}{$\overline{\log(\mstar/\msun)}$}&\multicolumn{4}{c}{$\overline{\log[\psi/(\msun \mathrm{yr}^{-1})]}$}	&\multicolumn{2}{c} {Number of galaxies}\\
&&&Low-mass$^{a}$&High-mass$^{b}$&&Low-mass$^{a}$&High-mass$^{b}$&&Low-mass$^{a}$&High-mass$^{b}$\\
\hline
&&&&&&&&&&\\
\vspace{0.1cm}
$0.20<z<0.45$	&\hb, \oiii, \ha&&$9.37(0.11)$&$10.48(0.13)$&&$-0.185(0.308)$&$0.375(0.347)$&&167&16\\
\vspace{0.1cm}
$0.45<z<0.70$	&\hb, \oiii&&$9.40(0.09)$&$10.56(0.12)$&&$-0.065(0.266)$&$0.765(0.387)$&&272&68\\
\vspace{0.1cm}
$0.70<z<0.85$	&\oii, \hb, \oiii&&$9.41(0.08)$&$10.54(0.11)$&	&\ \ \ $0.115(0.240)$&$0.855(0.396)$&&261&124\\
\vspace{0.1cm}
$0.85<z<1.10$	&\oii, \hb&&$9.43(0.09)$&$10.52(0.11)$&& \ \ \ $0.625(0.301)$&$1.155(0.546)$&&191&105\\
\vspace{0.1cm}
$1.10<z<1.40$	&\oii&&$9.50(0.10)$&$10.54(0.10)$	&&\ \ \ $0.965(0.229)$&$1.335(0.477)$&&96&121\\
\hline
\multicolumn{11}{l}{$^{a}$ Galaxies with stellar masses between 1.6 and $4.0 \times 10^{9}\msun$ (blue points in Figure~\ref{fig:msfr}).}\\
\multicolumn{11}{l}{$^{b}$ Galaxies with stellar masses between 2.5 and $6.3 \times 10^{10}\msun$ (green points in Figure~\ref{fig:msfr}).}\\
\end{tabular}
\end{small}
\end{center}
\end{table*}

\section{Star formation histories of blue galaxies}
\label{sec:results}

\begin{figure}
\begin{center}
\includegraphics[width=0.45\textwidth]{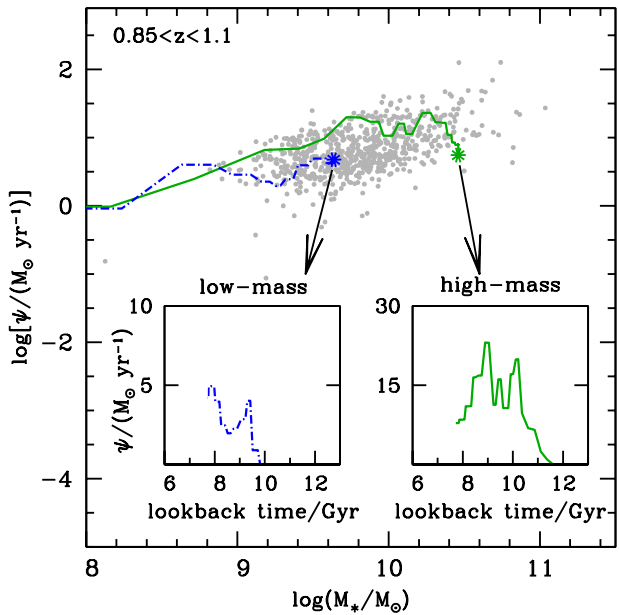}
\caption{Best estimates of SFR and stellar mass for the 900 galaxies at redshifts between 0.85 and 1.1 in the sample of \S\ref{sec:data} (grey points). The blue dash-dotted line represents the best-fit model SFH of an example galaxy of low stellar mass, while the green solid line represents the best-fit model SFH of an example galaxy of high stellar mass. Inset panels show SFR versus lookback time for these two example galaxies (at each time step, the SFR includes contributions from all components of the merger tree that end up in the galaxy by the redshift of observation). Both best-fit SFHs display significant fluctuations in time, and neither would be well approximated by a $\tau$-model.}
\label{fig:track}
\end{center}
\end{figure}

\begin{figure*}
\begin{center}
\includegraphics[width=\textwidth]{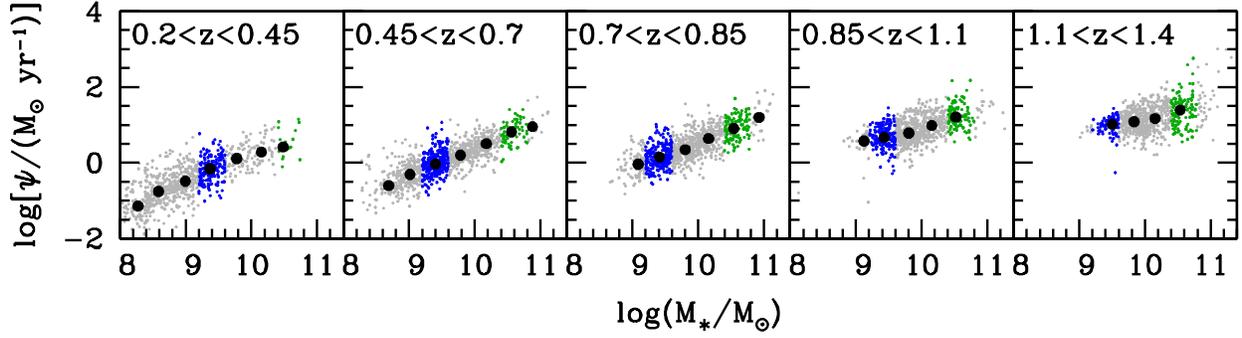}
\caption{Star-formation main sequence (i.e., SFR versus stellar mass) for the galaxies in five redshift bins spanning the range $0.2<z <1.4$ in the sample of \S\ref{sec:data}. A bin of low-mass galaxies ($\mstar = 1.6$--$4.0 \times 10^{9} \msun$) is highlighted in blue, and one of high-mass galaxies ($\mstar = 2.5$--$6.3 \times 10^{10} \msun$) highlighted in green.}
\label{fig:msfr}
\end{center}
\end{figure*}

\begin{figure*}
\begin{center}
\includegraphics[width=\textwidth]{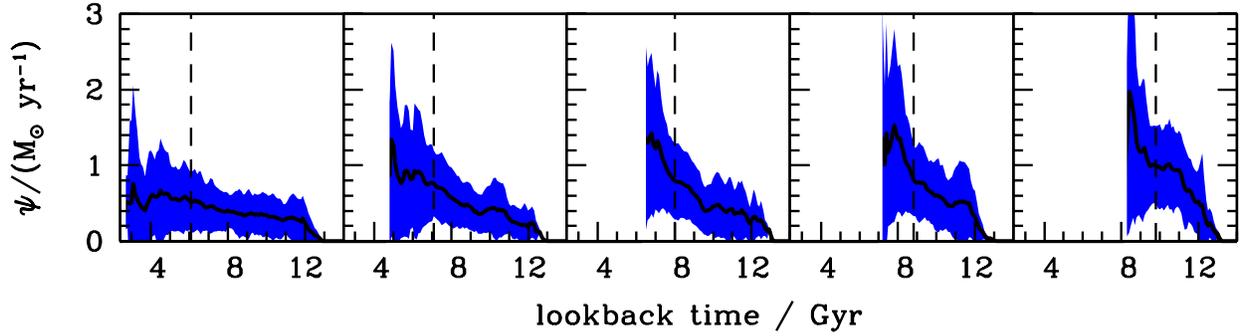}
\caption{Average best-fit SFHs of the galaxies in the low-mass bin of Figure~\ref{fig:msfr}, in the same redshift bins spanning the range $0.2<z<1.4$ from left to right (black solid line), and root mean square deviations about these averages (blue shadow). At any redshift, the average SFH of low-mass galaxies keeps rising in an extended way, inconsistent with declining $\tau$-models.}
\label{fig:sfhlow}
\end{center}
\end{figure*}

\begin{figure*}
\begin{center}
\includegraphics[width=\textwidth]{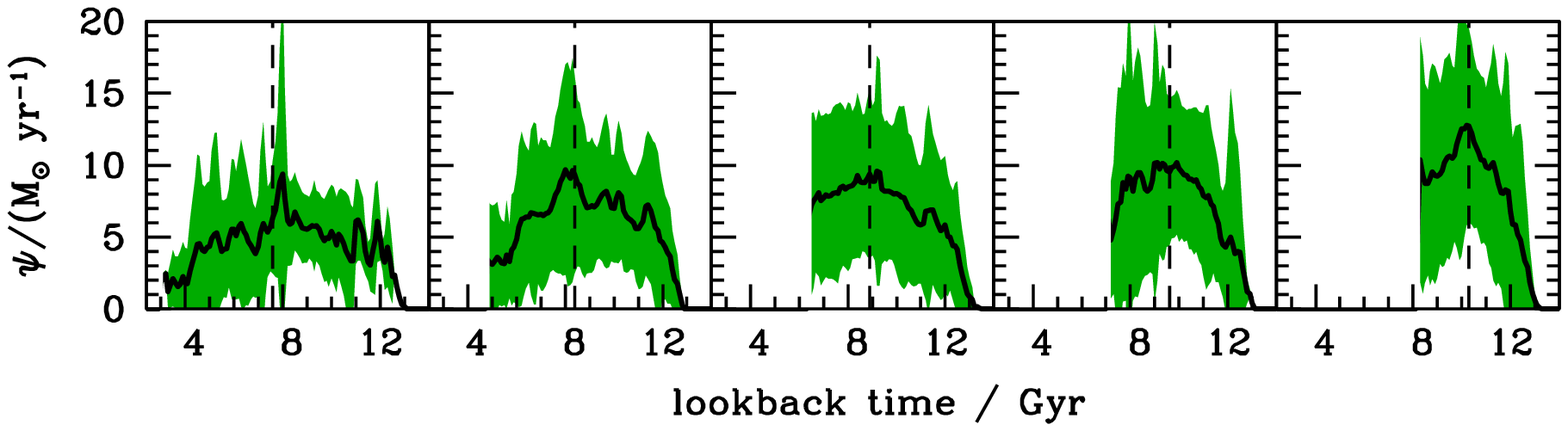}
\caption{Average best-fit SFHs of the galaxies in the high-mass bin of Figure~\ref{fig:msfr}, in the same redshift bins spanning the range $0.2<z<1.4$ from left to right (black solid line), and root mean square deviations about these averages (green shadow). At any redshift, the average SFH of high-mass galaxies rises and falls in a {\it bell-shaped} manner, which cannot be approximated by simple exponentially declining nor rising $\tau$-models. These findings imply that, at any redshift, low-mass galaxies (Figure~\ref{fig:sfhlow}) have on average higher current specific SFR ($\ssfr=\sfr/\mstar$) than high-mass galaxies.}
\label{fig:sfhhigh}
\end{center}
\end{figure*}

Figure~\ref{fig:track} illustrates the constraints derived on the star formation histories of blue galaxies at redshift around unity using the approach described in \S\ref{sec:model}.  We show the best estimates of \sfr\ as a function of \mstar\ for the 900 galaxies at redshifts between 0.85 and 1.1 in the sample of \S\ref{sec:data} (grey points). The results are consistent with the standard ``star-formation main sequence,''  which describes the systematic increase of the SFR with stellar mass and redshift (see e.g., \cite{brinchmann2004}, \cite{noeske2007}, \cite{daddi2007}, \cite{whitaker2012}).\footnote{\label{foot:SFMS} The normalization of the star-formation main sequence in Figure~\ref{fig:track} is about 0.2\,dex lower than that found by Noeske et al. (2007). This can be attributed to the different prescriptions adopted to derive SFR estimates and account for dust attenuation in that study.} Also shown in the inset panels in Figure~\ref{fig:track} are the best-fit SFHs of two example galaxies at similar redshift, $z\approx 0.9$, with similar {\it current} SFR,  $\psi\approx5 \msun \mathrm{yr}^{-1}$, but different stellar masses, $\mstar=4.3 \times 10^{9}\msun$ and $2.9 \times 10^{10}\msun$. It is interesting to note that the evolutionary paths of these galaxies remain within the star-formation main sequence over their entire lifetimes. The two SFHs show the characteristic fluctuations arising from the merger histories of the dark matter halos and the semi-analytic prescriptions used to model the baryons (gas infall, cooling, star formation, feedback). Neither SFH would be well approximated by a $\tau$-model.

We now explore the average SFHs of galaxies in different bins of stellar mass and redshift. Figure~\ref{fig:msfr} show the star-formation main sequence in five redshift bins spanning the range $0.2<z <1.4$. Along these relations, we select two stellar-mass bins sampled at all redshifts: a low-mass bin (1.6--$4.0\times 10^{9}\msun$) and a high-mass bin (2.5--$6.3 \times 10^{10}\msun$). Table~\ref{tab:data} lists the median stellar masses and SFRs in these two bins at all redshifts, together with the number of galaxies in each bin. For each redshift range, Figures~\ref{fig:sfhlow} and \ref{fig:sfhhigh} show the average best-fit SFHs of the galaxies in the low- and high-mass bins, respectively, as well as the root mean square deviations about these averages. In each case, the average SFH is computed by first recording the best-fit SFH of each galaxy at cosmic times prior to the redshift of observation, and then co-adding the individual SFHs of the different galaxies. The resulting SFH is normalized at each time step by dividing by the number of galaxies forming stars at that time. The wiggles in the SFHs of individual galaxies (see Figure~\ref{fig:track}) tend to be smoothed out by such averages, except in bins containing few galaxies (left-most panel in Figure~\ref{fig:sfhhigh}). We find that virtually identical average SFHs are obtained when selecting the second or third best-fit SFH for each galaxy.

The SFR of low-mass galaxies (Figure~\ref{fig:sfhlow}) increases steadily from early times to the epoch of observation, while that of high-mass galaxies (Figure~\ref{fig:sfhhigh}) first rises and then declines in a  {\it bell-shaped}  manner. Remarkably, at a fixed stellar mass, the shape of the average SFH is independent of galaxy redshift, although the evolution proceeds faster in the most distant galaxies. The lookback time at which galaxies had formed half their current stellar mass at a fixed redshift (vertical dashed lines in Figures~\ref{fig:sfhlow} and \ref{fig:sfhhigh}) increases with galaxy mass. This implies that the stars in massive galaxies in our sample are always typically older than those in low-mass galaxies. We note that the model library described in \S\ref{sec:lib} includes a comprehensive range of SFH shapes from declining, to rising, to bell-shaped, to erratically bursty at any redshift. Hence the results of Figures~\ref{fig:sfhlow} and \ref{fig:sfhhigh} do not simply reflect the prior distribution of SFHs in the model library. In fact, as shown by Figures~\ref{fig:sfhlow} and \ref{fig:sfhhigh} themselves, there is a significant dispersion in the SFHs of individual galaxies.

It is of interest to note that the average SFHs of the galaxies in our sample can be approximated by ``delayed $\tau$-models'' (\cite{bruzual1980}) of the type {\small $t^d\exp{(-t/\tau)}$}, with $\tau$ spanning the range from 0.2 to 1.0 (from high to low redshift) for the high-mass galaxies and from 0.4 to 3.0 for the low-mass ones. The exponent $d$ defines the ratio of the time of maximum SFR to the star formation timescale ($d=t_{\mathrm{max}}/\tau$). Of course, fluctuations arising from the specific merger history of any individual galaxy will make it deviate from the average SFH (Figure~\ref{fig:track}).

From the best-fit SFHs extracted for each galaxy in our analysis, we can also make predictions for the star-formation main sequence at earlier times. We find that at $z\approx 2$ the predicted dispersion about the median SFR in different bins of stellar mass along the sequence is roughly consistent with observations by e.g., Daddi et al. (2007), Reddy et al. (2012), and Whitaker et al. (2012). As expected from the star-formation main sequence we observe at $z<1.4$ (see footnote~\ref{foot:SFMS}), the normalization of the sequence is slightly lower than in these other studies.

\section{Summary and conclusion}
\label{sec:concl}

We have measured the SFHs of 4517 blue galaxies at redshifts between 0.2 and 1.4 in the AEGIS Survey using a new, sophisticated spectral analysis tool. To achieve this, we built a comprehensive library of galaxy SEDs by combining a semi-analytic post-treatment of the Millennium cosmological simulation (\citep{springel2005}) with state-of-the-art models of stellar population synthesis, nebular emission, and attenuation by dust. For each galaxy in our sample, we retrieved the best-fit model SFH and the probability density functions of stellar mass and SFR using a Bayesian analysis. We reach the following conclusions:
\begin{itemize}
\item The best-fit SFHs of individual galaxies exhibit characteristic fluctuations arising from the merger histories, which cannot be well described by simple $\tau$-models.
\item At all redshifts in the explored range $0.2< z< 1.4$, the average SFH of high-mass blue galaxies ($\sim4\times10^{10}\msun$) rises and falls in a {\it bell-shaped} manner, suggesting that these galaxies are gradually turning off their star formation activity. This behavior cannot be approximated by simple exponentially declining nor rising $\tau$-models.
\item The average SFH of low-mass galaxies ( $\sim3\times10^9\msun$) keeps rising in a more extended way, also inconsistent with declining $\tau$-models.
\item The average SFHs of both high- and low-mass galaxies can be reasonably well approximated by delayed $\tau$-models of the form {\small $t^d\exp(-t/\tau)$}.
\item The shapes of the SFHs suggest that, at any redshift, low-mass galaxies have on average higher specific SFR ($\ssfr=\sfr/\mstar$) than high-mass galaxies, consistent with the trends in the star-formation main sequence (e.g., \cite{noeske2007}) and in the latest simulations of Behroozi et al. (2012).
\item At fixed stellar mass, the shape of the average SFH is independent of galaxy redshift, although the evolution proceeds faster in the most distant galaxies.  
\item At fixed redshift, the lookback time by which half the stellar mass of a galaxy has formed (vertical dashed lines in Figures~\ref{fig:sfhlow} and \ref{fig:sfhhigh}) increases with stellar mass.
\item The SFHs retrieved from our analysis tend to make the galaxies remain within the star-formation main sequence for their entire lifetimes.
\end{itemize}

The results presented in this Letter will be extended in the near future by exploiting optical stellar absorption-line signatures and infrared photometry to improve the constraints on the SFHs (Pacifici et al. in preparation).

\acknowledgments

This work was supported in part by the KASI-Yonsei Joint Research Program (2012) for the Frontiers of Astronomy and Space Science funded by the Korea Astronomy and Space Science Institute, and in part by the Marie Curie Initial Training Network ELIXIR of the European Commission under contract PITN-GA-2008-214227. C.P. thanks the JWST Project and the Astrophysics Science Division at Goddard Space Flight Center for hosting her while working on this paper, and Andrea Macci\`o and Aaron Dutton for useful discussions. S.A.K is supported by an appointment to the NASA Postdoctoral Program at NASA's Goddard Space Flight Center, administered by Oak Ridge Associated Universities through a contract with NASA. The authors also acknowledge NSF grants AST 95-29098 and 00-71198 to UC Santa Cruz. This study makes use of data from AEGIS, a survey conducted with the Chandra, GALEX, Hubble, Keck, CFHT, MMT, Subaru, Palomar, Spitzer, VLA, and other telescopes and supported in part by the NSF, NASA, and the STFC.

\def\aj{AJ}
\def\araa{ARA\&A}
\def\apj{ApJ}
\def\apjl{ApJ}
\def\apjs{ApJS}
\def\apss{Ap\&SS}
\def\aap{A\&A}
\def\aapr{A\&A~Rev.}
\def\aaps{A\&AS}
\def\mnras{MNRAS}
\def\pasp{PASP}
\def\pasj{PASJ}
\def\qjras{QJRAS}
\def\nat{Nature}

\def\aplett{Astrophys.~Lett.}
\def\aas{AAS}
\let\astap=\aap
\let\apjlett=\apjl
\let\apjsupp=\apjs
\let\applopt=\ao



\begin{thebibliography}{}

\bibitem[Behroozi et al. 2012]{behroozi2012} {Behroozi}, P.~S., {Wechsler}, R.~H., \& {Conroy}, C. 2012, arXiv:1207.6105

\bibitem[Bell \& de Jong 2001]{bell2001} Bell, E.~F., \& de Jong, R.~S. 2001, \apj, 550, 212

\bibitem[Brinchmann et al. 2004]{brinchmann2004} {Brinchmann}, J., {Charlot}, S., {White}, S.~D.~M., {et al.} 2004, \mnras, 351, 1151

\bibitem[Bruzual \& Charlot 2003]{bruzual2003} Bruzual, A.~G., \& Charlot, S. 2003, \mnras, 344, 1000

\bibitem[Bruzual \& Kron 1980]{bruzual1980} Bruzual, A.~G. \& Kron, R.~G. 1980, \apj, 241, 25

\bibitem[Bundy et al. 2006]{bundy2006} {Bundy}, K., {Ellis}, R.~S., {Conselice}, C.~J., et al. 2006, \apj, 651, 120

\bibitem[Chabrier 2003]{chabrier2003} {Chabrier}, G. 2003, \pasp, 115, 763

\bibitem[Charlot \& Fall 2000]{charlot2000} {Charlot}, S., \& {Fall}, S.~M. 2000, \apj, 539, 718

\bibitem[Charlot \& Longhetti 2001]{charlot2001} Charlot, S., \& Longhetti, M. 2001, \mnras, 323, 889

\bibitem[Coil et al. 2004]{coil2004} {Coil}, A.~L., {Newman}, J.~A., {Kaiser}, N., et al. 2004, \apj, 617, 765

\bibitem[Daddi et al. 2007]{daddi2007} {Daddi}, E., {Dickinson}, M., {Morrison}, G., {et al.} 2007, \apj, 670, 156

\bibitem[Davis et al. 2007]{davis2007} {Davis}, M., {Guhathakurta}, P., {Konidaris}, N.~P., {et al.} 2007, \apjl, 660, L1

\bibitem[De Lucia \& Blaizot 2007]{delucia2007} {De Lucia}, G., \& {Blaizot}, J. 2007, \mnras, 375, 2

\bibitem[Donley et al. 2007]{donley2007} {Donley}, J.~L., {Rieke}, G.~H., {P{\'e}rez-Gonz{\'a}lez}, P.~G., {et al.} 2007, \apj, 660, 167

\bibitem[Ferland 1996]{ferland1996} {Ferland}, G.~J. 1996, Hazy, A Brief Introduction to Cloudy 90, Internal Report, Univ. Kentucky, Lexington

\bibitem[Kauffmann et al. 2003]{kauffmann2003} {Kauffmann}, G., {Heckman}, T.~M., {Tremonti}, C., {et al.} 2003, \mnras, 346, 1055

\bibitem[Klypin et al. 2011]{klypin2011} {Klypin}, A.~A., {Trujillo-Gomez}, S., \& {Primack}, J. 2011, \apj, 740, 102

\bibitem[Lee et al. 2009]{lee2009} {Lee}, S.-K., {Idzi}, R., {Ferguson}, H.~C., {et al.} 2009, \apjs, 184, 100

\bibitem[Maraston et al. 2010]{maraston2010} {Maraston}, C., {Pforr}, J., {Renzini}, A., {et al.} 2010, \mnras, 407, 830

\bibitem[Newman et al. 2012]{newman2012} {Newman}, J.~A., {Cooper}, M.~C., {Davis}, M., {et al.} 2012, arXiv:1203.3192

\bibitem[Noeske et al. 2007]{noeske2007} {Noeske}, K.~G., {Weiner}, B.~J.,  {Faber}, S.~M., {et al.} 2007, \apjl, 660, L43

\bibitem[Pacifici et al. 2012]{pacifici2012} {Pacifici}, C., {Charlot}, S., {Blaizot}, J., \& {Brinchmann}, J. 2012, \mnras, 421, 2002

\bibitem[Papovich et al. 2001]{papovich2001} {Papovich}, C., {Dickinson}, M., \& {Ferguson}, H.~C. 2001, \apj, 559, 620

\bibitem[Pforr et al. 2012]{pforr2012} {Pforr}, J., {Maraston}, C., \& {Tonini}, C. 2012, \mnras, 422, 3285

\bibitem[Reddy et al. 2012]{reddy2012} {Reddy}, N.~A., \& {Pettini}, M., \& {Steidel}, C.~C., {et al.} 2012, \apj, 754, 25

\bibitem[Shapley et al. 2005]{shapley2005} {Shapley}, A.~E., \& {Steidel}, C.~C., \& {Erb}, D.~K., {et al.} 2005, \apj, 626, 698

\bibitem[Springel et al. 2005]{springel2005} {Springel}, V., \& {White}, S.~D.~M., \& {Jenkins}, A., {et al.} 2005, \nat, 435, 629

\bibitem[Stark et al. 2009]{stark2009} {Stark}, D.~P., {Ellis}, R.~S., {Bunker}, A., {et al.} 2009, \apj, 697, 1493

\bibitem[Weiner et al. 2007]{weiner2007} {Weiner}, B.~J., {Papovich}, C., {Bundy}, K. {et al.} 2007, \apjl, 660, L39

\bibitem[Whitaker et al. 2012]{whitaker2012} {Whitaker}, K.~E., {van Dokkum}, P.~G., {Brammer}, G., \& {Franx}, M. 2012, \apjl, 754, L29

\bibitem[Wuyts et al. 2009]{wuyts2009} {Wuyts}, S., {Franx}, M., {Cox}, T.~J., {et al.} 2009, \apj, 696, 348

\bibitem[Yan et al. 2011]{yan2011} {Yan}, R., {Ho}, L.~C., {Newman}, J.~A., {et al.} 2011, \apj, 728, 38

\end{thebibliography}
\end{document}